\documentclass[aps,showpacs,twocolumn,superscriptaddress]{revtex4}
\usepackage{epsf}
\usepackage{graphicx}
\usepackage{dcolumn}

\newcommand{\tresj}[6]{\left( \begin{array}{ccc}
                              #1 & #2 & #3 \\
                              #4 & #5 & #6 
                             \end{array}
                      \right)}

\newcommand{\nk}{{\bf k}}
\newcommand{\np}{{\bf p}}
\newcommand{\nq}{{\bf q}}
\newcommand{\nr}{{\bf r}}

\newcommand{\nl}{\mbox{\boldmath{$l$}}}

\newcommand{\nA}{{\bf A}}

\newcommand{\nJ}{{\bf J}}

\newcommand{\nY}{{\bf Y}}

\newcommand{\nV}{{\bf V}}

\newcommand{\hr}{{\bf \hat{r}}}

\newcommand{\nsigma}{\mbox{\boldmath$\sigma$}}

\begin{document}

\title{Equivalence between local Fermi gas and 
       shell models in  inclusive muon capture from nuclei}


\author{ J.E. Amaro}
\affiliation{Departamento de F\'{\i}sica Moderna, 
          Universidad de Granada,
          Granada 18071, Spain}
\author{C. Maieron}
\affiliation{INFN, Sezione di Catania,
              Via Santa Sofia 64, 
              95123 Catania, Italy}
\author{J. Nieves}
\affiliation{Departamento de F\'{\i}sica Moderna, 
          Universidad de Granada,
          Granada 18071, Spain}
\author{M. Valverde}
\affiliation{Departamento de F\'{\i}sica Moderna, 
          Universidad de Granada,
          Granada 18071, Spain}  

\begin{abstract}

Motivated by recent studies of inclusive neutrino nucleus processes
and muon capture within a correlated local Fermi gas model (LFG),
we discuss the relevance of nuclear finite size effects in these
reactions at low energy, in particular for muon capture.  To
disentangle these effects from others coming from the reaction
dynamics we  employ here a simple uncorrelated shell model
that embodies the typical finite size content of the problem.
The integrated decay widths of muon atoms calculated with this shell
model are then compared for several nuclei 
with those obtained within the uncorrelated
LFG, using in both models exactly the same theoretical ingredients and
parameters.  We find that the two predictions are in quite
good agreement, within 1--7\%, when the shell model density and the correct
energy balance is used as input
in the LFG calculation.  The present study indicates that, despite the
low excitation energies involved in the reaction, integrated inclusive
observables, like the total muon capture width, are quite independent of the
fine details of the nuclear wave functions.

\end{abstract}



\maketitle


\section{Introduction}


In this paper we study the importance of nuclear finite size effects
in inclusive muon capture reactions. 
The motivation for this investigation comes from 
the results of a recent publication \cite{Nie04,Nie04b}, where we
have developed a model which describes rather well the inclusive
$^{12}$C$(\nu_\mu,\mu^-)$ and $^{12}$C$(\nu_e,e^-)$ cross sections
near threshold, and inclusive muon capture by nuclei.
This approach, which is an extension of the quasi-elastic
inclusive electron scattering model of \cite{Gil97}, is based on a
Local Fermi Gas (LFG), where the simplicity of the model makes it
possible to include a great variety of effects into the reaction
dynamics \cite{Ch90,Muk98,Sin93,Kos96,Sin98}.  In particular
long-range nuclear correlations are taken into account by computing
the RPA polarization propagator containing nucleonic and
$\Delta(1232)$ degrees of freedom.  All these effects
are crucial for the correct
analysis of atmospheric neutrino fluxes \cite{Nuint04,Fu98, Zuber} and
to describe the recent neutrino experiments
\cite{Kra92,Bod94,Aue01,Albert95,Atha97,Auer02}.

The results of \cite{Nie04}, particularly those of muon
capture, indicate that for some kind of inclusive reactions
the validity of the LFG can be extended to
energies lower than expected. Although the LFG leads to reasonable
predictions for integrated quantities, at low energies
it is not possible to describe the shape of the
differential neutrino cross section or muon capture width within this
model.  In fact both the discrete and the continuum states of the
final nucleus, including giant resonances, contribute.  
However, when one sums over all the final states, the information
about the fine details of the spectrum is lost and only the global
contribution remains.  Therefore for some integrated inclusive
observables, the results depend mainly on global quantities such as
the correct energy balance or the nucleon distribution.  An example is
the inclusive pion capture model of \cite{Ama97,Chi90}.
It is also worth to mention the  pioneering work of 
ref. \cite{Bar96} 
in the context of comparing shell model with Fermi gas
for $u$-inclusive neutrino scattering.

The goal of this paper is to investigate whether finite nucleus
effects can affect significantly the LFG results of ref. \cite{Nie04}
for inclusive muon capture.  From the present calculation it is
possible to estimate an uncertainty of 1--7\% due to finite size
effects not taken into account in the LFG calculation of \cite{Nie04}.
There already exist microscopic calculations of neutrino-nucleus
reactions and muon capture, based on the RPA or large shell model (SM)
basis \cite{Mea01,Ko94,Ko00,Kol99,Hay00,Vol00,Krm04}, and other
approaches such as the relativistic shell model \cite{Mai03} and the
Green's function method \cite{Meu03}.  All of them treat correctly the
finite size of the system.  However from direct comparison of these
models with the LFG it is not possible to deduce the importance of the
finite nucleus treatment.  The reason is that such comparison should
be done between models embodying exactly the same nuclear
dynamics. Instead a great variety of residual interactions, shell
model wave functions, current operators, Coulomb effects, etc., have
been used in these works, making it impossible to 
disentangle the impact of the finite size on the different contributions.

Before such comparison be undertaken in a sophisticated framework, it
is convenient to understand the uncorrelated case.  For this reason we
have chosen to perform this comparison with a simple model where
the finite-size effects can be easily recognized.  Thus here we consider
the extreme SM, i.e., single particle states in a Woods-Saxon (WS)
potential, and we compare the results with those obtained from
a LFG model.  We do not include
long range correlations of RPA type or configuration mixing.  In
addition we use the static form of the single-nucleon charged current
(CC), in order to simplify even more the calculation and to draw cleaner
conclusions.
On the other hand, within our simplified approach, we do not confine 
ourselves to the single case of $^{12}$C, but make a more thorough
study of how finite size effects vary with 
increasing nuclear mass, by comparing the results for a set of
closed shell nuclei: $^{12}$C, $^{16}$O, $^{40}$Ca, and $^{208}$Pb.

The present calculation represents a first-stage test case to
check the ``equivalence'' of LFG and shell models for some inclusive
processes.  We choose the $\mu$-capture reaction for this
investigation since it involves low excitation energies,
the worst conditions for the LFG. 
Of course, under these simplifications it
makes no sense to compare our results with the experimental data
\cite{Suz87}, nor it is the intention of this work, since it was
already done in ref. \cite{Nie04} with the full model.

\section{General Formalism}

\subsection{Partial and differential width}

Here we present the formalism to describe the inclusive muon capture 
within our model. We use Bjorken\& Drell \cite{Bjo65}
conventions.
We consider a negative muon bound into an initial 
 nucleus $^A_Z X$, which decay 
into a final nucleus $^A_{Z-1}Y$ plus a muon neutrino (not detected).
\begin{equation}  \label{decay}
\mu^- + ^A_Z X \longrightarrow \, ^A_{Z-1}Y + \nu_\mu 
\end{equation}
The final nucleus can be in the discrete or in the continuum.
We assume that the initial muon is in a $s$-wave state
(normalized to one)
\begin{equation}
   \phi_{1s}(\nr)= \phi_{1s}(r)= \frac{R_{1s}(r)}{\sqrt{4\pi}}.
\end{equation}
We describe the wave function $\phi_{1s}$ in a non-relativistic
framework by solving the Schr\"odinger equation for the muon in the nuclear 
Coulomb potential, including finite size and vacuum polarization effects.    
The final neutrino has four-momentum $k'^{\mu}=(\epsilon',\nk')$.
The leptonic current matrix element involved in the decay is then
\begin{equation} \label{lepton-current}
\langle \nu_\mu|j^\mu(x)|\mu\rangle =
\ell^{\mu} 
\phi_{1s}(x)  
e^{ik'\cdot x}
\end{equation}
where $x^\mu=(t,\nr)$ is the space-time coordinate,
 $\phi_{1s}(x)$ is the time-depending muon wave function 
\begin{equation}
\phi_{1s}(x) = \phi_{1s}(r) e^{-i\epsilon t}
\end{equation}
and $\epsilon$ is its initial energy (including the binding).  Since
we treat the muon as non-relativistic, we describe its spin by a Pauli
spinor $\chi$ which is contained into the leptonic vector 
$\ell^\mu$, defined by 
\begin{equation} \label{lepton-vector}
\ell^\mu = \left[\frac{m'}{V\epsilon'}\right]^{1/2}
\overline{u}_\nu(\nk')\gamma^\mu(1-\gamma_5)u_\mu(0)
\end{equation}
where we have written the muon
four-spinor as $u_\mu (0) = (\chi, 0)$, i.e., corresponding
effectively to a four-spinor with momentum zero. This is equivalent to
neglect in the following the initial muon momentum $\nk=0$ in the kinematics
---however the full spatial dependence of the wave function $\phi_{1s}(\nr)$
is maintained in the matrix element, see below.
Finally in (\ref{lepton-vector}) $V$ is the normalization volume of the 
neutrino plane wave, and $m'$ its mass, that is set to zero at the end 
of the calculation. 

The $S$-matrix element relevant for the decay reaction (\ref{decay}) is then
\begin{equation}
S_{fi} = -2\pi i \delta(E_f-E_i-\omega) \frac{G}{\sqrt{2}} {\ell}^\mu
\langle f | \tilde{J}_\mu(-\nk')|i\rangle
\end{equation}
where $|i\rangle$ and $|f\rangle$ are the initial and final nuclear states,
with energies $E_i$ and $E_f$ respectively, $\omega=\epsilon-\epsilon'$ is the
energy transfer, $G=1.1664\times 10^{-5} \rm GeV^{-2} \cos\theta_c$ is the
Fermi coupling constant multiplied by the cosine of Cabibbo's angle, and we
have introduced the effective current operator $\tilde{J}_\mu$, defined in
coordinate space as
\begin{equation}
\tilde{J}_\mu(\nr) = J_\mu(\nr)\phi_{1s}(r).
\end{equation}
Here $J_\mu(\nr)$ is the nuclear CC operator to be specified
below.  Finally, $\tilde{J}_\mu(\nq)$ is the Fourier transform
\begin{equation}
\tilde{J}_\mu(\nq) = \int d^3r e^{i\nq\cdot\nr} \tilde{J}_\mu(\nr).
\label{Fourier}
\end{equation}

The differential decay width can be computed easily. Since the initial
nucleus is unpolarized, the distribution of neutrinos is
independent of the angles, and the corresponding 
angular integral gives a factor $4\pi$.

At this point we have to distinguish two cases, depending on the 
kind of final state $|f\rangle$ reached. In the first case the final nucleus 
is in a discrete state, that can be the ground state or an excited state.
The neutrino energy takes discrete values
 fixed by energy conservation,  
and the partial width for the transition 
$i\rightarrow f$ is written as, 
\begin{equation}
\Gamma_{i\rightarrow f} = 
 \frac{G^2}{2\pi}\frac{\epsilon'}{m}\eta^{\mu\nu}
W^{i\rightarrow f}_{\mu\nu}(q).
\end{equation}
where $q=|\nq|$, and the usual leptonic tensor 
has been introduced
\begin{equation}
\eta^{\mu\nu}=
k^\mu k'^\nu
+k^\nu k'^\mu
-m\epsilon' g^{\mu\nu}
+i \epsilon^{\mu\nu\alpha\beta}k_\alpha k'_\beta
\end{equation}
for initial muon momentum $k^\mu = (m,0)$, where $m$ is the muon mass.
We have also defined the muon-hadronic tensor for the transition
\begin{equation} \label{hadronic}
W^{i\rightarrow f}_{\mu\nu}(q)= \overline{\sum_{M_fM_i}}
\langle f | \tilde{J}_\mu(\nq)|i\rangle^*
\langle f | \tilde{J}_\nu(\nq)|i\rangle.
\end{equation}
where $\nq=\nk-\nk'=-\nk'$ is the momentum transfer,
  we sum over final spin components $M_f$, 
and average over  initial spins $M_i$.

In the second case, the final nucleus goes to the continuum, above the
one-particle emission threshold, and the final neutrino energy ranges between
0 and the maximum energy available minus the nucleon separation energy of the
final nucleus. The continuum spectrum of neutrinos is described by the
differential decay width
\begin{equation}
\frac{d\Gamma_c}{d\epsilon'}
= \frac{G^2}{2\pi}\frac{\epsilon'}{m}\eta^{\mu\nu}W^{(c)}_{\mu\nu}(q,\omega).
\end{equation}
where now the continuum hadronic tensor is defined as
\begin{equation} \label{hadronic-c}
W^{(c)}_{\mu\nu}(q,\omega)= \overline{\sum_{fi}}\delta(E_f-E_i-\omega)
\langle f | \tilde{J}_\mu(\nq)|i\rangle^*
\langle f | \tilde{J}_\nu(\nq)|i\rangle.
\end{equation}
Here a sum over final (continuum) states and an average over initial spin
is assumed. 

The contraction between the leptonic and muon-hadronic tensor is 
easily performed in a coordinate system where the $z$ axis is 
in the $\nq$ direction.
We finally obtain the following expression for the differential decay width
\begin{equation}
\frac{d\Gamma_c}{d\epsilon'}
= \frac{G^2}{2\pi}\epsilon'^2 
\left( R_{C}+R_L-2R_{CL}+R_T+2R_{T'} \right)
\end{equation}
and a similar expression for the discrete partial widths, 
where for simplicity the response functions have been introduced
as the following components of the hadronic tensor \cite{Don79,Ama05}:
\begin{eqnarray}
R_{C} &=&W^{00} \label{RC}\\
R_{CL} &=& -\frac12\left( W^{03}+W^{30} \right)\\
R_L  &=& W^{33}  \\
R_T &=&  W^{11}+W^{22} \\
R_{T'} &=& -\frac{i}{2}\left( W^{12}-W^{21} \right) \label{RTP}
\end{eqnarray}

The total (inclusive) width is obtained by integrating and summing 
over the continuum and discrete, respectively
\begin{equation}
\Gamma = \sum_f \Gamma_{i\rightarrow f} 
+ \int_0^{\epsilon'_{\rm max}} \frac{d\Gamma_c}{d\epsilon'} d \epsilon'
\end{equation}

\subsection{Multipole expansion}

Since the shell model states have good angular momentum,
$|i\rangle=|J_iM_i\rangle$, $|f\rangle=|J_fM_f\rangle$, it is usual to
perform analytically the sums over third components using the
Wigner-Eckart theorem.  To this end one begins with the following
multipole expansion valid for the components of any current operator
in momentum space as a sum of operators with good angular momentum of
rank $J$ (note that the $z$-axis is in the $\nq$ direction)
\begin{eqnarray}
\tilde{J}_0(q) &=& \sqrt{4\pi} \sum_{J=0}^{\infty}i^J [J] \hat{C}_{J0}(q)
\label{rhoex}
\\
\tilde{J}_z(q) &=& -\sqrt{4\pi} \sum_{J=0}^{\infty}i^J [J] \hat{L}_{J0}(q)
\label{Jzex}
\\
\tilde{J}_m(q) &=& -\sqrt{2\pi} \sum_{J=0}^{\infty}i^J [J] 
             \left[\hat{E}_{Jm}+m \hat{M}_{Jm}(q)\right], \medskip m=\pm1
\nonumber \\
\label{Jmex}
\end{eqnarray}
where we use the notation $[J]\equiv\sqrt{2J+1}$, and
in the last equation the spherical components of the current 
vector have been introduced
$J_{\pm 1}=\mp(J_x\pm J_y)/\sqrt{2}$.
The operators in this expansion are the usual Coulomb, longitudinal, 
transverse electric and transverse magnetic operators, defined by
\begin{eqnarray}
\hat{C}_{J0}(q) &=& 
\int d^3 r j_{J}(qr) Y_{J0}(\hr)\tilde{J}_0(\nr)
\\
\hat{L}_{J0}(q) &=& 
\frac{i}{q}\int d^3 r \nabla\left[j_{J}(qr) Y_{J0}(\hr)\right]
\cdot\tilde{\nJ}(\nr)  
\label{LJoperator}\\
\hat{E}_{Jm}(q) &=& 
\frac{1}{q}\int d^3 r 
\nabla\times\left[j_{J}(qr)\nY_{JJm}(\hr)\right]
\cdot\tilde{\nJ}(\nr)
\label{EJoperator}\\
\hat{M}_{Jm}(q) &=& 
\int d^3 r j_{J}(qr) \nY_{JJm}(\hr)\cdot\tilde{\nJ}(\nr)
\label{MJoperator}
\end{eqnarray}
where $j_J$ is a spherical Bessel Function and $\nY_{JJm}$ is a vector
spherical harmonic. Note that the above expansions
(\ref{rhoex}--\ref{Jmex}) are a direct consequence of the familiar
plane wave expansion in spherical Bessel functions and spherical
harmonics, inside the Fourier transform (\ref{Fourier}).

Inserting the expansions (\ref{rhoex}--\ref{Jmex}) inside the 
hadronic tensor (\ref{hadronic-c}) we obtain
\begin{eqnarray}
R_C 
&=& 
\frac{4\pi}{2J_i+1} \sum_{J}
|C_J|^2
\\
R_L
&=& 
\frac{4\pi}{2J_i+1} \sum_{J}|L_J|^2
\\
R_{CL}
&=& 
\frac{2\pi}{2J_i+1} \sum_{J} \left(C_J^*L_J+L_J^*C_J \right)
\\
R_T
&=& 
\frac{4\pi}{2J_i+1} \sum_{J}
\left(|E_J|^2+|M_J|^2\right)
\\
R_{T'}
&=& 
-\frac{2\pi}{2J_i+1} \sum_{J} \left(E_J^*M_J+M_J^*E_J \right)
\label{RT'}
\end{eqnarray}
for the responses in the discrete, and a similar expression for the
continuum responses with the addition of a sum over final states and a
delta of energies $\sum_f \delta(E_f-E_i-\omega)$. 
The multipole coefficients in these sums are the reduced matrix elements
of the corresponding multipole operators
\begin{eqnarray}
C_J(q) &=& \langle f \| \hat{C}_J(q)\| i\rangle
\\
L_J(q) &=& \langle f \| \hat{L}_J(q)\| i\rangle
\\
E_J(q) &=& \langle f \| \hat{E}_J(q)\| i\rangle
\\
M_J(q) &=& \langle f \| \hat{M}_J(q)\| i\rangle
\end{eqnarray}
 The values of $J$
and $J_f$ are related by angular momentum conservation $|J_i-J_f| \leq
J \leq J_i+J_f$. In the particular case of closed-shell nuclei, such as
$^{12}$C, with $J_i=0$, we have $J_f=J$.

\subsection{Weak charged current}

In order to simplify the comparison with the LFG, in this first stage
we apply the above formalism to the CC $J^\mu=V^\mu-A^\mu$ in
the static limit.  This is not unreasonable for the $\mu$-capture
reaction since all the momenta involved are small. Thus we only
maintain the leading order in the standard expansion of the matrix
element of the vector current
\begin{equation}
V^\mu(\np',\np)=
\overline{u}(\np') 
\left[ 2F_1^V \gamma^\mu + i\frac{2F_2^V}{2M}\sigma^{\mu\nu}Q_\nu
\right]
u(\np)
\end{equation}
in powers of $p/M$, $p'/M$, with $M$ the nucleon mass, and 
$Q^\mu=(\omega,\nq)$ the four-momentum transfer ($Q^2=\omega^2-q^2$).
Therefore we take
\begin{eqnarray}
V^0 &\simeq&  2F_1^V \\
\nV & \simeq& 0.
\end{eqnarray}
In the case of the axial current
\begin{equation}
A^\mu(\np',\np)=
\overline{u}(\np') 
\left[ G_A \gamma^\mu\gamma^5 + G_PQ^\mu\gamma^5
\right]
u(\np)
\end{equation}
we expand taking into account that, from PCAC, the pseudo-scalar form 
factor $G_P$ is of order $O(M)$
\begin{equation}
G_P=\frac{2M}{m_\pi^2-Q^2}G_A
\end{equation}
and the leading-order term in the expansion of the axial current becomes
\begin{eqnarray}
A^0 & \simeq & -\frac{G_A}{m_\pi^2-Q^2} (\nq\cdot\nsigma) \omega\\
\nA & \simeq & G_A\nsigma-\frac{G_A}{m_\pi^2-Q^2}(\nq\cdot\nsigma)\nq.
\end{eqnarray}
Therefore the total weak CC in the static limit that we use in the
present work is
\begin{eqnarray}
J^0 & = & J_V^0 - J_P^0 \label{J0} \\
\nJ & = & -\nJ_A-\nJ_P  \label{Jvec}
\end{eqnarray}
and the different terms in these equations are defined below.
First order terms in an expansion in powers of $1/M$ not included in 
our calculation can give an appreciable contribution, but the present 
approximation is enough for our purposes of testing the equivalence between 
LFG and shell models.

\subsection{Multipole matrix elements of the current}

The different multipoles of the  vector, axial and
pseudo-scalar currents ($J_V^0$, $\nJ_A$ and $J_P^\mu$), introduced in
eqs.~(\ref{J0},\ref{Jvec}), are computed following the
approach of ref. \cite{Ama96}, where the matrix elements of the
electro-weak neutral current were considered in the context of
parity-violating electron scattering.  In the case of the vector
current we only consider the zeroth component to leading order
$J_V^0=2F_1^V$. Therefore only the Coulomb multipoles of this current
enter in our calculation. The reduced matrix elements between single
nucleon wave functions, with angular momentum quantum numbers
$(l_p,j_p)$ and $(l_h,j_h)$, are given by
\begin{equation}
\langle p \| \hat{C}_J(q)\| h \rangle
= 2F_1^V P^+_{l_p+l_h+J} [J]\, a_J I_J(q).
\end{equation}
Here we use the notation $P^+_n$ for the parity function 
(=1 if $n$ is even and 0 if $n$ is odd), and we 
have defined the function $I_J(q)$
\begin{equation}
I_J(q) = \int_0^\infty dr\, r^2 j_J(qr) R_p^*(r)R_h(r) \phi_\mu(r)
\end{equation}
which contains the dynamical information on the nuclear transition and
the muon wave function.  Finally the coupling coefficient $a_J$ is defined
in terms of a three-j coefficient
\begin{equation}
a_J\equiv \frac{(-1)^{j_p+1/2}[j_p][j_h]}{\sqrt{4\pi}}
\tresj{j_p}{j_h}{J}{\frac12}{-\frac12}{0}.
\end{equation}
In the case of the axial current we only consider the space components
$\nJ_A = G_A \nsigma$ (we neglect the time component to leading order)
so only the longitudinal and transverse (electric and magnetic)
matrix elements enter:
\begin{eqnarray}
\langle p \| \hat{L}^A_J(q)\| h \rangle
&=& iG_AP^+_{l_p+l_h+J+1}
\frac{a_J}{[J]}
\nonumber\\
&&
\mbox{}\times
\big[ 
(\kappa_p+\kappa_h-J) I_{J-1}(q)
\nonumber\\
&&\mbox{}+(\kappa_p+\kappa_h+J+1) I_{J+1}(q)
\big]
\\
\langle p \| \hat{E}^A_J(q)\| h \rangle
&=& -iG_AP^+_{l_p+l_h+J+1}
\frac{a_J}{\sqrt{J(J+1)}[J]}
\nonumber\\
&&
\mbox{}\times
\big[ 
(J+1+\kappa_p+\kappa_h) J I_{J+1}(q)
\nonumber\\
&&
\mbox{}+(J-\kappa_p-\kappa_h) (J+1) I_{J-1}(q)
\big]
\\
\langle p \| \hat{M}^A_J(q)\| h \rangle
&=&
 G_AP^+_{l_p+l_h+J}
\frac{a_J[J]}{\sqrt{J(J+1)}}
\nonumber\\
&&
\mbox{}\times
(\kappa_p-\kappa_h) I_{J}(q),
\end{eqnarray}
where we use the notation $\kappa_p= (-1)^{j_p+l_p+\frac12}(j_p+\frac12)$.
Note that the longitudinal and electric multipoles have 
abnormal parity, i.e., $l_p+l_h+J= \mbox{odd}$  as expected 
for an axial current.

In the case of the pseudo-scalar current $J_P^\mu=
-\frac{G_A}{m_\pi^2-Q^2}(\nq\cdot\nsigma)Q^\mu$, the multipoles can be
related to the longitudinal components of the axial current $J_A^z =
G_A \nsigma\cdot\hat{\nq}$.  Using the expansion (\ref{Jzex}) for the
longitudinal current we have for the zero-th component
\begin{eqnarray}
J_P^0 
&=& -\frac{\omega q}{m_\pi^2-Q^2} J_A^z \\
&=& \frac{\omega q}{m_\pi^2-Q^2} \sqrt{4\pi}\sum_J i^J [J]\hat{L}_{J0}^A.
\end{eqnarray}
Comparing with the expansion (\ref{rhoex}) we obtain that the Coulomb
operators of the pseudo-scalar current are proportional to the
longitudinal multipoles of the axial current,
\begin{equation}
\hat{C}_{J0}^P = \frac{\omega q}{m_\pi^2-Q^2}\hat{L}_{J0}^A
\end{equation}
and the same relation holds for the matrix elements.
Since the spatial part of the pseudo-scalar current is proportional to $\nq$, 
it has no transverse components. Only the  longitudinal multipoles
enter, that are again proportional to the axial ones
\begin{equation}
\hat{L}_{J0}^P = -\frac{q^2}{m_\pi^2-Q^2}\hat{L}_{J0}^A.
\end{equation}
and a similar relation between  the corresponding matrix elements.

Finally, note that in the present static approximation, where there are no
transverse multipoles for the vector current, the response
function $R_{T'}=0$, because only the interference between electric and
magnetic multipoles of the vector and axial current, respectively,
(and vice-versa) would enter in eq. (\ref{RT'}).

\subsection{The local Fermi gas}

In the local Fermi gas model we first compute the decay width 
$\Gamma_{FG}[\rho_P,\rho_N]$ 
for a muon at rest inside a Fermi gas with constant proton and neutron
densities 
\begin{equation} \label{rho}
\rho_P=k_{FP}^3/3\pi^2, 
\kern 1cm
\rho_N=k_{FN}^3/3\pi^2,
\end{equation}
where $k_{FP}$ and $k_{FN}$ are the Fermi momenta of protons and neutrons,
respectively.
 With the charged current (\ref{J0},\ref{Jvec}), the response
functions (\ref{RC}--\ref{RTP}) are computed in this model using the formalism
of \cite{Nie04,Ama96}. The final result can be written simply as
\begin{equation}
\frac{d\Gamma_{FG}}{d\epsilon'}
= \frac{G^2}{\pi}\epsilon'{}^2
\left[ 4F_{1V}^2+G_A^2(3+C_P^2-2C_P)
\right] R_0
\end{equation}
where we have defined the following factor coming from the pseudo-scalar 
current
\begin{equation}
C_P\equiv\frac{m \epsilon'}{m_\pi^2-Q^2},
\end{equation}
and the function $R_0$ is related to the imaginary part of the Linhard
function \cite{Nie04,Ama96}
\begin{equation}
2R_0 = -\frac{1}{\pi}{\rm Im}\,\overline{U}
=
\frac{M^2}{2\pi^2q}\theta(\epsilon_{FP}-\epsilon_0)(\epsilon_{FP}-\epsilon_0).
\end{equation}
Here we have defined
\begin{equation}
\epsilon_0 = 
{\rm Max}
\left\{
\epsilon_{FN}-\omega,\frac{1}{2M}\left(\frac{M\omega}{q}-\frac{q}{2}\right)^2
\right\}
\end{equation}
and $\epsilon_{FP}= k_{FP}^2/2M$ is the Fermi energy of protons, 
and $\epsilon_{FN}= k_{FN}^2/2M$ for neutrons.
The LFG width is then obtained by inserting the proton and neutron 
densities, $\rho_P(\nr)$ and $\rho_N(\nr)$, 
of the finite size nucleus into eq. (\ref{rho}) 
and averaging with the muon density \cite{Nie04}
\begin{equation}
\Gamma_{LFG}=\int d^3 r |\phi_\mu(\nr)|^2 \Gamma_{FG}[\rho_P(\nr),\rho_N(\nr)].
\end{equation}
An important input for 
 the LFG is the experimental $Q$-value
for the reaction (\ref{decay})
\begin{equation} \label{Q-value}
Q= M( ^A_{Z-1}Y )-M(^A_Z X) = \omega_{\rm min}
\end{equation}
which is the minimum value allowed for the energy transfer $\omega$.  In order
to account for this value in the Fermi gas, we substitute $\omega$ by
$\omega-Q$, since part of the energy $\omega$ 
is employed in producing the final
nucleus. In this way we treat correctly the energy balance, which is
important for describing the experimental muon-capture width \cite{Nie04}.
When different densities are used for protons and neutrons, especially 
in the case of $^{208}$Pb, there is a gap 
\begin{equation}\label{gap}
\epsilon_{\rm gap}=\epsilon_{FN}-\epsilon_{FP}
\end{equation}
between neutron and proton Fermi energies, that has to be considered also in
the energy balance by substituting 
\begin{equation} \label{omega-gap}
\omega \longrightarrow \omega+\epsilon_{\rm gap}-Q.
\end{equation}

\section{results}

In this section we present results for a set of closed-shell nuclei 
$^{12}$C, $^{16}$O, $^{40}$Ca, and $^{208}$Pb.
In the extreme shell model the initial and final nuclear 
wave function are described as Slater determinants
constructed  with single-particle wave functions that are solutions of the
 Schr\"odinger equation with a Woods-Saxon potential
\begin{equation} \label{potencial}
V(r)=V_0 f(r,R_0,a_0)
-V_{LS}\frac{2\nl\cdot\nsigma}{r}\frac{df(r,R_0,a_0)}{dr}+V_C(r)
\end{equation}
where
\begin{equation}
f(r,R_0,a_0)=\frac{1}{1+e^{(r-R_0)/a_0}}
\end{equation}
and $V_C(r)$ is, for protons,  
the Coulomb potential of a charged sphere of charge $Z-1$
and radius $R_C$, and it is equal to zero for neutrons.
The parameters of the potential are commonly  fitted  to 
the experimental energies of the valence shells or the 
charge radius. In the present case of muon capture 
we fit the experimental $Q$-value (\ref{Q-value}) 
for the decay reaction (\ref{decay}).
In the shell model, the energy difference between hadronic final and initial
states is computed as the difference between the corresponding shells
\begin{equation}
\omega = \epsilon_p -\epsilon_h
\end{equation}
where $\epsilon_p$ and $\epsilon_h$ are eigenvalues of the Schr\"odinger
equation for particles (neutrons) and holes (protons), respectively.
Therefore the $Q$ value (\ref{Q-value}) is obtained in this model as the 
energy difference between the first unoccupied neutron shell and the 
last occupied proton shell, corresponding to the transition 
of a valence proton to a neutron above the Fermi level.
This makes only one condition for fixing the several parameters of the 
potential (\ref{potencial}). 
Wherever possible we set the remaining  parameters of
the potential to values similar to the ones used in other studies
like those of refs. \cite{Ama96,Ama97,Alb03}.
In our calculation we use different sets of 
parameters, denoted WS1, WS2 and WS3,
shown in table \ref{tableI} for
$^{12}$C, $^{16}$O, $^{40}$Ca, and in table \ref{tableII} for $^{208}$Pb.

\begin{table}
\begin{tabular}{ccrrrrrr}
        &    & $V^P_0$ & $V^P_{LS}$ & $V^N_0$ & $V^N_{LS}$ & $r_0$& $a_0$\\
\hline\hline
$^{12}$C&WS1 & $-52.38$ & $-20.30$ & $-50.85$ & $-24.11$  & 1.25 & 0.57 \\
        &WS2 & $-62.38$ &  $-3.20$ & $-50.85$ & $-18.40$  &      &      \\
        &WS3 & $-62.38$ &  $-3.20$ & $-38.30$ & $ -3.15$  &      &      \\
\hline
$^{16}$O&WS1 & $-52.50$ & $-0.60$  & $-52.50$ & $-0.60$   & 1.27 & 0.53 \\
        &WS2 & $-52.50$ & $-7.00$  & $-42.80$ & $-6.54$   &      &      \\
        &WS3 & $-50.00$ & $0.00$   & $-50.00$ & $0.00$    &      &      \\
\hline
$^{40}$Ca&WS1& $-50.45$ & $-4.83$  & $-48.66$ & $-5.20$   & 1.25 & 0.53 \\
        &WS2 & $-57.50$ & $-11.11$ & $-55.00$ & $-2.30$   &      &      \\
        &WS3 & $-57.50$ & $-11.11$ & $-53.00$ & $-5.10$   &      &      \\
\hline\hline
\end{tabular}
\caption{\label{tableI}
Parameters of the Woods-Saxon potentials used in 
$^{12}$C, $^{16}$O and  $^{40}$Ca for protons (P) and
neutrons (N). The units are MeV for $V_i$, and fm
for $a_0$ and $r_0$. The reduced radius parameter $r_0$ 
is defined by
$R_0=r_0 A^{1/3}$. The Coulomb radius is chosen $R_C=R_0$.}
\end{table}

\begin{table}
\begin{tabular}{ccrrrrrr}
    &   & $V_0$   & $V_{LS}$ & $r_0$ & $a_0$ &  $r_{LS}$ & $a_{LS}$ \\
\hline\hline
WS1 & P & $-60.4$ & $-7.45$  & 1.26  & 0.79  &  1.21     & 0.59       \\
    & N & $-46.9$ & $-5.64$  & 1.21  & 0.66  &  1.17     & 0.64       \\
WS2 & P & $-60.4$ & $-6.75$  & 1.26  & 0.79  &  1.22     & 0.59       \\
    & N & $-43.5$ & $-6.08$  & 1.26  & 0.66  &  1.17     & 0.64      \\
\hline\hline
\end{tabular}
\caption{\label{tableII} Parameters of the Woods-Saxon potentials of
$^{208}$Pb for protons (P) and neutrons (N). Note that we use
different radius parameters for the central and spin-orbit parts of
the potential. The units are MeV for $V_i$, and fm for $a_i$ and
$r_i$. The reduced radius parameters $r_i$ are defined by $R_i=r_i
A^{1/3}$. The Coulomb radius is chosen $R_C=R_0$.}
\end{table}

The only states relevant for $\mu$-capture are the occupied proton holes and
the neutron particles above the valence shell. In the discrete sector several
transitions are possible with fixed excitation energies.  The single particle
energies of the last occupied proton shell and first unoccupied neutron shell
obtained with the potentials of tables \ref{tableI},\ref{tableII} are shown in
table \ref{tableIII}.  The $Q$ value corresponds to the transition
$P\longrightarrow N$ in table \ref{tableIII}, with an energy difference
\begin{equation} \label{Qvalue}
Q=\epsilon(N)-\epsilon(P)
\end{equation}
which is also shown in table \ref{tableIII}, 
together with the experimental value $Q_{\rm  exp}$ in the last column.
The number of
discrete neutron states is finite.  Above the last discrete neutron state, the
next allowed transitions are to the continuum.  The continuum neutron states
are obtained by solving the Schr\"odinger equation for positive energies. More
details on the continuum solutions can be found in refs.
\cite{Ama96,Ama97,Alb03}.

\begin{table}
\begin{tabular}{ccrrrr}
Nucleus    &             &    WS1 &   WS2  &   WS3  & esp  \\\hline\hline
$^{12}$C   & $P1p_{3/2}$ &$-15.96$&$-18.38$&$-18.13$&      \\
           & $N1p_{1/2}$ & $-2.08$& $-4.50$& $-4.25$&      \\
           & $Q$-value   &  13.88 &  13.88 &  13.88 & 13.880\\\hline
$^{16}$O   & $P1p_{1/2}$ &$-15.31$&$-12.77$&$-13.76$&       \\
           & $N1d_{5/2}$ &$ -4.39$&$ -1.84$&$ -2.83$&       \\
           & $Q$-value   &  10.92 &  10.93 &  10.93 & 10.931\\\hline
$^{40}$Ca  & $P1d_{3/2}$ &$ -8.33$&$ -8.78$&$ -8.78$&       \\
           & $N1f_{7/2}$ &$ -6.51$&$ -6.95$&$ -6.95$&        \\
           & $Q$-value   &   1.83 &   1.83 &   1.83 &  1.822 \\\hline
$^{208}$Pb & $P3s_{1/2}$ &$ -8.19$&$ -8.19$&        &         \\ 
           & $N2g_{9/2}$ &$ -2.68$&$ -2.68$&        &         \\
           & $Q$-value   &   5.51 &   5.51 &        &  5.512 \\ \hline\hline
\end{tabular}
\caption{\label{tableIII} Single particle energies in MeV 
used in the fit of the $Q$-value for $\mu$-capture
(the experimental values are shown in the last column).}
\end{table}

In order to compare with the LFG, it is important to use 
as input the proton and neutron densities obtained in the corresponding 
shell model,
by summing  over occupied states as follows:  
\begin{equation}
\rho_P(\nr)=\sum_{\rm protons} \frac{2j+1}{4\pi}|R_{nlj}(r)|^2,
\end{equation}
where $R_{nlj}(r)$ are the radial wave functions,
and a similar expression for neutrons.

\begin{table}[t]
\begin{tabular}{cccccr}
         &     & discrete  &  total  & LFG     &     \%  \\ \hline\hline
$^{12}$C & WS1 &   0.3115  & 0.4406  & 0.4548  &    3.2  \\ 
         & WS2 &   0.3179  & 0.4289  & 0.4360  &    1.7  \\ 
         & WS3 &   0.2746  & 0.5510  & 0.4732  & $-14.1$ \\ \hline
$^{16}$O & WS1 &   1.113   & 1.282   & 1.360   &    6.1  \\ 
         & WS2 &   0.590   & 1.118   & 1.392   &   24.3  \\ 
         & WS3 &   1.154   & 1.332   & 1.387   &    4.1  \\ \hline
$^{40}$Ca& WS1 &  29.10    & 37.12   & 36.73   &  $-1.1$ \\
         & WS2 &  27.79    & 33.79   & 34.90   &    3.3  \\
         & WS3 &  26.28    & 32.73   & 35.03   &    7.0  \\\hline
$^{208}$Pb&WS1 &  215.6    & 390.3   & 399.4   &    2.3  \\
          &WS2 &  266.8    & 467.4   & 439.5   &   -5.9  \\ \hline\hline
\end{tabular}
\caption{\label{tableIV} Integrated width in units of $10^{5} s^{-1}$ 
for the different nuclei and Woods-Saxon potentials, compared with the 
LFG results using the corresponding charge densities. 
The discrete contribution of the shell model is shown in the first column. }
\end{table}

\begin{figure}[tb]
\begin{center}
\includegraphics[scale=0.7, bb=150 370 399 786]{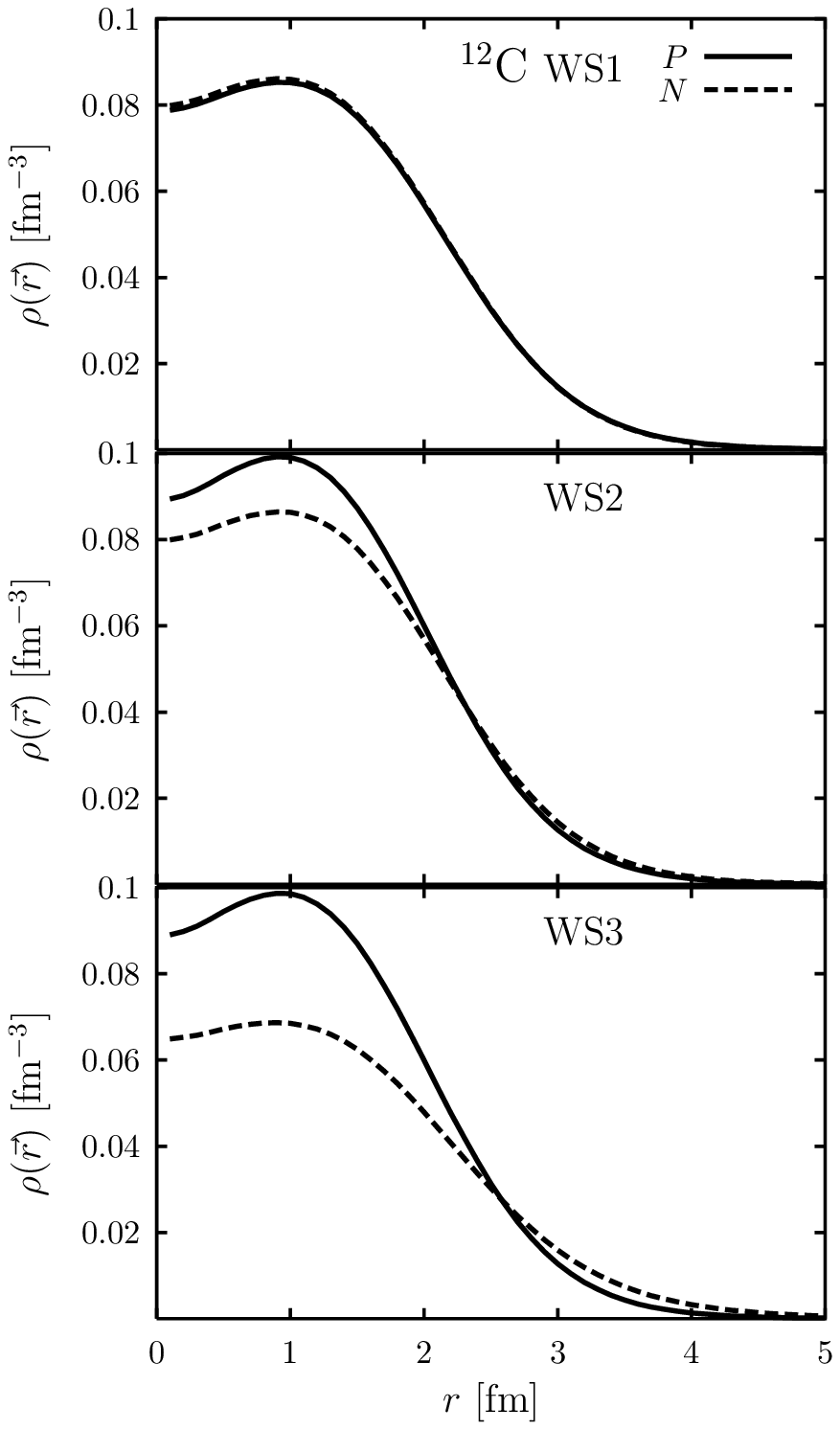}
\end{center}
\caption{Proton and neutron densities of $^{12}$C 
for the several WS potentials used in this work.
}
\end{figure}

The shell model calculation has been checked by comparison with the
factorized plane wave impulse approximation (PWIA) \cite{Ama96b}.  In
this approximation there is no final-state interaction and hence the
final neutron states are plane waves. The transition matrix elements
appearing in the hadronic tensor (\ref{hadronic-c}) are computed
trivially in terms of the product of a single nucleon current matrix
element times the Fourier transform of a nuclear overlap function of
the missing momentum, in a way which is similar to the analysis of
exclusive $(e,e'p)$ reactions \cite{Maz02} (but this time the nuclear
overlap function includes the bound muon wave function).  As a
consequence, the exclusive hadronic tensor factorizes as the product
of a single nucleon hadronic tensor times a partial momentum
distribution, and the calculation is straightforward  in the shell model.  For
the present case an additional integration and a sum over initial
states is needed since we are interested in the inclusive case,
similar to the factorized PWIA in $(e,e')$ introduced in
\cite{Ama96b}.  The PWIA can be also approached with our multipole
expansion code by setting to zero the WS potential in the final
states.  This allows us to check the multipole expansion calculation
and, at the same time, to fix the number of multipoles in the sum over
$J$, eqs. (\ref{rhoex}--\ref{Jmex}).  The differences with the
factorized calculations are negligible when we include up to five
multipoles.

\begin{figure}[tb]
\begin{center}
\includegraphics[scale=0.7, bb=150 280 450 780]{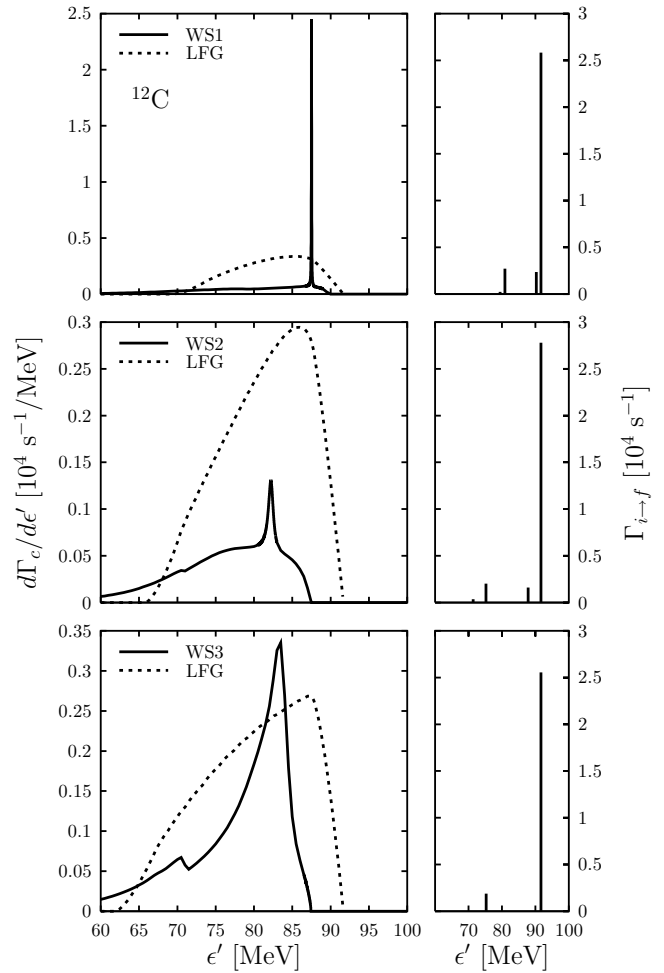}
\end{center}
\caption{Differential SM width of $^{12}$C 
to the continuum (left panels) compared to the
  LFG, and partial widths to the discrete states (right panels), as a function
  of the neutrino energy, for the different WS potentials considered in this
  work.  }
\end{figure}

In table \ref{tableIV} we show results for the integrated inclusive widths for
the four nuclei and for the different models used in this work.  For each one
of the WS parameterizations we show in the second column the contribution to
the width from the discrete final neutron states, while in the third column we
show the total width (discrete + continuum). The LFG results are shown in the
fourth column, and for comparison we show the percentual relative difference
between LFG and WS in the last column.  Next we discuss the results obtained
for each one of the nuclei studied in this work.

\subsection{$^{12}$C.}
In table \ref{tableIV}
we can see that, in the case of WS1 and WS2, the LFG and WS results for
$^{12}$C are 
quite similar, differing only in $\sim 2$--3\%. In the case of WS3 the
differences are larger, around 14\%.

\begin{figure}[tb]
\begin{center}
\includegraphics[scale=0.7, bb=150 370 399 786]{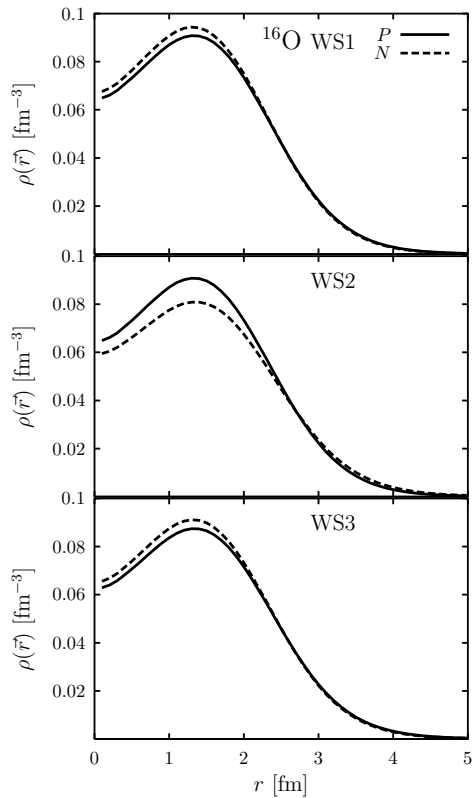}
\end{center}
\caption{The same as Fig. 1 for 
$^{16}$O.
}
\end{figure}

All the WS potential parameters have been fitted to the experimental $Q$-value
(\ref{Qvalue}), which, in the SM, is the difference between the neutron
$p_{1/2}$ and proton $p_{3/2}$ energies.  Among all the potentials, WS1 is the
more realistic since it also fits the proton and neutron separation energies
of $^{12}$C and give reasonable masses for the ground states of the $^{13}$N
and $^{13}$C nuclei.  When we use similar parameters for protons and neutrons,
like in WS1, we need a large spin-orbit splitting in order to fit the
experimental $Q$-value.  In the case of the potential WS2 we use different
parameters for protons and neutrons: The proton well is similar to the one of
ref. \cite{Ama96}, that is more attractive than WS1, with small spin orbit
strength.  The neutron parameters are similar to WS1.
 Finally in WS3 we have used a small neutron
spin-orbit splitting, as for protons, but we had to make the neutron well much
less attractive than for protons. Apart from changing the single-particle
energies, the effect of modifying the WS potential can be appreciated in the
proton and neutron densities shown in Fig.~1.  For more attractive potentials
the nucleus becomes more dense in the interior.  For this reason, the WS3
neutron density turns out to be the smallest one, while the proton density is
around 3/2 the neutron one.  Hence the LFG results are worse for very
different neutron and proton densities. In this situation the proton and
neutron Fermi momenta (\ref{rho}) are clearly different, leading to a gap
between proton and neutron energies, (\ref{gap}), which in this case is
negative, since the density is smaller for neutrons, and
$\epsilon_{FN}<\epsilon_{FP}$.  Therefore a proton near the Fermi surface can
decay to a neutron above the neutron Fermi surface with an energy decrement.
This is an unrealistic situation, since precisely in this case the neutrons
are less bound than protons in the SM, and therefore lie at higher energies.
Another argument to disregard this case is the well known property of closed
(sub)shell nuclei such as $^{12}$C, for which the neutron and proton densities
should be similar. Note that in all cases the gap between the $N$ and $P$ 
Fermi species has been taken into account in the energy balance 
by the replacement (\ref{omega-gap}).

\begin{figure}[tb]
\begin{center}
\includegraphics[scale=0.7, bb=150 280 450 780]{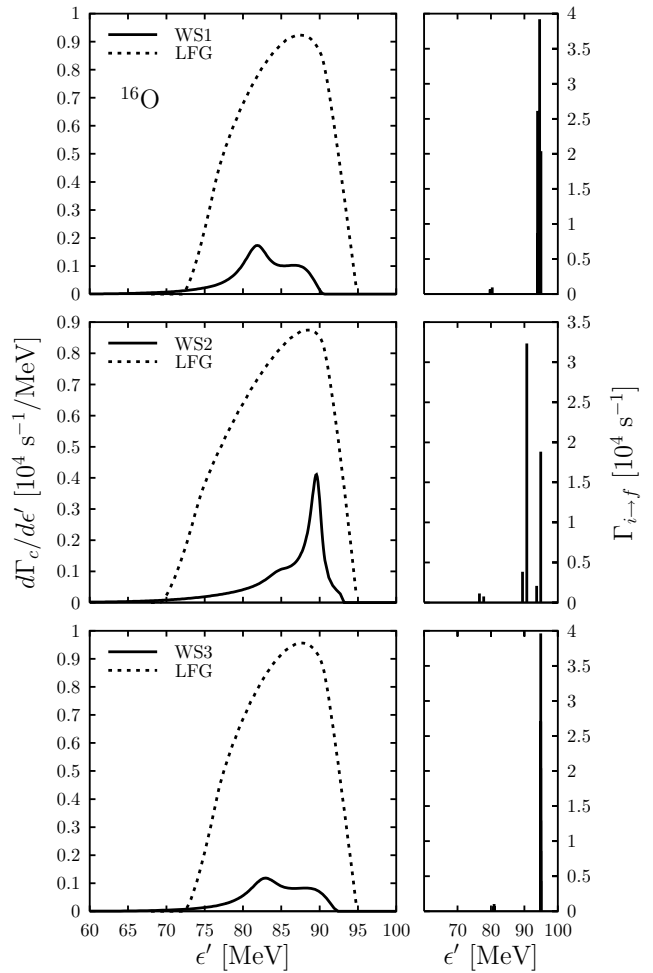}
\end{center}
\caption{The same as Fig. 2 for $^{16}$O.}
\end{figure}

In Fig.2 we compare the SM results for the differential width to the continuum
with the LFG distribution for the different WS parameters (left panels). The
shapes of both distributions are completely different.  The differences are
more apparent for the WS1 potential, where there is a very high and sharp
neutron resonance in the SM spectrum.  The partial widths to the discrete
states are shown in the right panels of Fig.~2.  Considering these differences
in shape between the LFG and the SM, it is a very notable result that the
integrated widths (adding the discrete states) take similar values  in both
models as was shown in table \ref{tableIV}. 
This outcome agrees with the findings of ref. \cite{Cen97},
where the same problem was addressed in the context of inelastic electron 
scattering on nuclei.

 The biggest contribution to the
width comes in all the cases from the transition to the ground state, and its
magnitude does not depend very much on the potential, since in the transition
$p_{3/2}\rightarrow p_{1/2}$ the wave functions in the initial and final
states are similar across the different potentials. Note also that there are
transitions to final discrete states that lie in the continuum (particularly,
transitions from the $1s$ shell). These states will contribute to the giant
resonances after an appropriate treatment of the residual interaction (such as
in the RPA). Under the light of the present preliminary study and the results
of ref. \cite{Nie04} one expects that the inclusion of the RPA does not change
too much our conclusions and the total integrated width be similar for
correlated LFG and SM.

\begin{figure}[tb]
\begin{center}
\includegraphics[scale=0.7, bb=150 370 399 786]{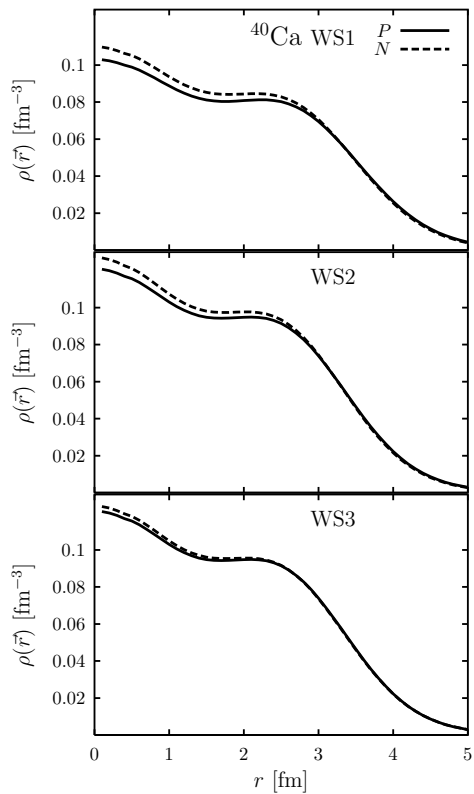}
\end{center}
\caption{The same as Fig. 1 for 
$^{40}$Ca.
}
\end{figure}

\begin{figure}[tb]
\begin{center}
\includegraphics[scale=0.7, bb=150 280 450 780]{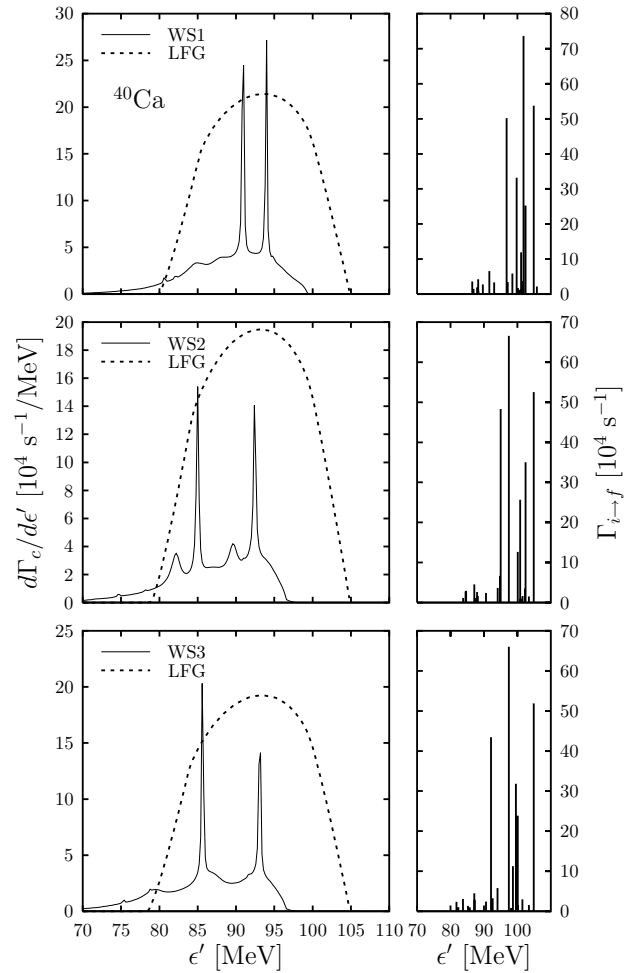}
\end{center}
\caption{The same as Fig. 2 for $^{40}$Ca.}
\end{figure}

\subsection{$^{16}$O}

In the case of $^{16}$O the integrated widths computed in the LFG are also
very close, $\sim 4$--6\%, to the SM results with the potentials WS1 and WS3
(see table \ref{tableIV}).  The worse results are obtained for the WS2
parameterization; the corresponding width is 24\% of the SM one.  This can
also been understood in terms of what was said for the case of $^{12}$C above,
by looking at the $^{16}$O densities shown in Fig. 3.  The case of WS2 is the
only one where the protons are more bound than the neutrons, hence the
$N$-density is smaller than the $P$-density inside the nucleus, which is again
an unrealistic situation because one expect the opposite in a closed shell
nucleus such as $^{16}$O.

We should add that the  $^{16}$O nucleus is delicate in the sense that 
 the experimental
$Q$-value of 10.93 MeV is too large to be fitted by the WS parameters found in
the literature \cite{Ama97,Ama96}. In fact in the SM the $Q$-value is the
difference between the $N1d_{5/2}$ and $P1p_{1/2}$ energies (see table
\ref{tableIII}). The effect of the spin-orbit potential is to increase
$\epsilon(p_{1/2})$ and to decrease $\epsilon(d_{5/2})$, that is, goes to
reduce the $Q$-value. (The opposite happens for $^{12}$C, where the 
$Q$-value is the difference between the $Np_{1/2}$ and $Pp_{3/2}$ energies.
Hence the spin-orbit goes to increase the $Q$-value.) 
Therefore to make that value as large as 11 MeV one
needs a small spin-orbit potential, as in WS1, or to rise the neutron well
with respect to the proton well, as in WS2, at the cost of making the neutrons
less bound than protons. The first option is preferred because it allows for
similar proton and neutron densities.  Precisely the third parameterization
WS3 has been chosen with $V_{LS}=0$ to maximize the difference between these
two shells.

\begin{figure}[tb]
\begin{center}
\includegraphics[scale=0.7, bb=150 500 399 786]{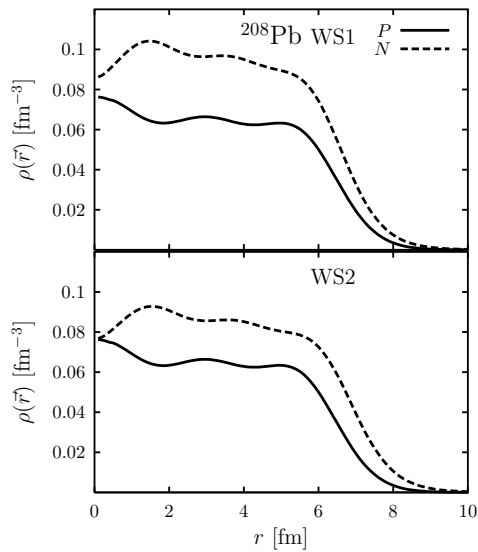}
\end{center}
\caption{The same as Fig. 1 for 
$^{208}$Pb.
}
\end{figure}

The differential and partial widths of $^{16}$O for the three WS potentials
are shown in Fig. 4. In the three cases the 
main contribution comes from the discrete spectrum
(see also table \ref{tableIV}). Since $V_{LS}=0$ for WS3,
 the dominant contribution comes from
transitions from the $1p$ to the $1d$ shell.

\subsection{$^{40}$Ca}

The LFG results improve when the mass of the nucleus increases as in the
present case
of the nucleus $^{40}$Ca. In fact, from table \ref{tableIV} we see that for
this nucleus the LFG integrated width is within 1\% of the SM result for WS1,
and 3\% and 7\% in the other two cases. This improvement was expected because
the Fermi gas description of the nucleus should work better for heavier
nuclei. In the case of WS2 and WS3 the proton parameters have been fixed to
the typical values used in the literature, and we have fitted the neutron
ones.  Since here the experimental $Q$-value is small, $Q=1.8$ MeV,
one does not need to change too much the typical neutron parameters. 
In the case of WS1 we have tried to maintain the $P$ and $N$
parameters similar. The
proton and neutron densities are close in all cases, as shown in Fig. 5, and
the proton levels always lie above the neutron ones.
 
The neutrino spectrum shown in Fig. 6 presents a more complex
structure than the lighter nuclei discussed above. More potential
resonances arise and the discrete spectrum presents more lines
distributed along the allowed energy region.

\subsection{$^{208}$Pb}

Finally we discuss the results for the closed-shell heavy nucleus $^{208}$Pb.
In table \ref{tableIV} we present integrated widths only for two sets of
potential parameters, WS1 and WS2. This is the only case where we use
different radius parameters for protons and neutrons and also for the central
and spin-orbit parts of the potential, see table \ref{tableII}.
In both cases the LFG results are close,
within 3 and 6\%, to the SM ones.  The $Q$ value, 5.5 MeV, is close to the
$N2g_{9/2}$ and $P3s_{1/2}$ energy difference of typical parameterizations
\cite{Ama96,Deh77}. Only small variations of these parameterizations found in
the literature are allowed if one wants to maintain the ordering of the
energies around the Fermi level.  Also only small variations are needed to fit
the experimental $Q$-value.  In the present case the treatment of asymmetric
nuclear mater is essential, because the proton and neutron densities, shown in
Fig. 7, are clearly different. Therefore the correct treatment of the gap in
the energy balance, eqs. (\ref{gap},\ref{omega-gap}), is needed to obtain the
results of table \ref{tableIV}. 
Moreover in this case the neutrino spectrum shown in Fig. 8 shows also 
an improved resemblance between the  LFG and SM (although numerous potential
resonances appear), even taking into account
the distribution of the discrete spectrum.   

\begin{figure}[tb]
\begin{center}
  \includegraphics[scale=0.7, bb=150 435 450 780]{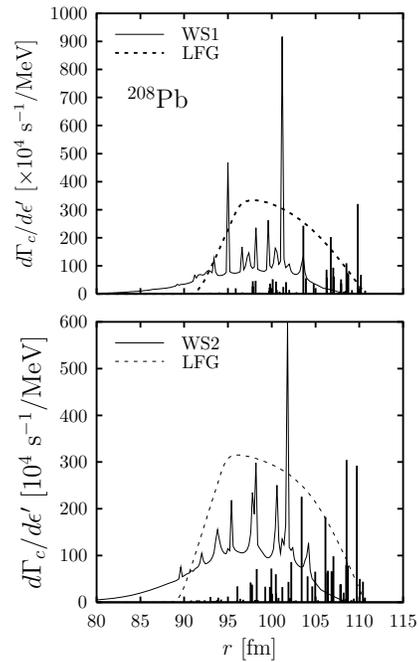}
\end{center}
\caption{The same as Fig. 2 for $^{208}$Pb. In this case we show the discrete 
contributions in units of $10^4 s^{-1}$ in the same panel as the continuum
one.
}
\end{figure}

\section{Conclusions}
 
In this paper we have estimated the magnitude of the finite nucleus effects on
inclusive muon capture, aiming at quantifying the uncertainty of the LFG
results of ref. \cite{Nie04}. It is not possible to disentangle these effects
by comparison with the highly sophisticated RPA or shell models existing in
the literature due to the different theoretical ingredients embodied in them.

To know how much the LFG is modified by finite size effects, one would need a
finite nucleus model with exactly the same input as the LFG, in order to make
the comparison meaningful.  Obviously this would be a drawback because
precisely one wants to use the LFG due to its simplicity, in order to include
very complex dynamical effects hard to incorporate in finite nucleus
treatments.  Therefore before using a very sophisticated model, it is
convenient to see what happens in the uncorrelated case.

In this paper we have focused on a simple shell model without nuclear
correlations, but that contains the relevant information about the finite
nuclear structure, and we have compared it with the uncorrelated LFG using the
same input.  In particular the SM proton and neutron densities have been used
in the LFG calculation.  We have applied both models to a set of closed-shell
nuclei: $^{12}$C, $^{16}$O, $^{40}$Ca, and $^{208}$Pb.  In the SM we fit the
experimental $Q$-value of the decay, while the same value is used to correct
the energy transfer in the LFG, taking into account also the gap between
neutron and proton Fermi energies.  As expected, the neutrino spectrum is very
different in both models, in particular the LFG cannot account for the
resonances and discrete states.  However, in the case of the lighter nuclei,
$^{12}$C and $^{16}$O, the SM and LFG results for the integrated width are
close ---within 3--6\%--- for WS parameters with similar neutron and proton
densities, but the results are somewhat different, within 14--24\%, for the
disregarded cases in which the protons lie below the neutrons.  For the medium
and heavy nuclei, $^{40}$Ca and $^{208}$Pb, the integrated widths are always
very close, within 1--7\%.  The final neutrino spectra of the LFG becomes more
similar to the SM, including the discrete part, for heavier nuclei.  Under the
assumption that RPA correlations and finite size effects are somewhat
decoupled for integrated inclusive observables, the present results can
explain why the LFG results of ref.  \cite{Nie04} describe so well the
experimental data.

\section*{Acknowledgments}

This work was partially supported by funds provided by DGI (Spain) and
FEDER funds, under Contract No. BFM2002-03218, 
and by the Junta de Andaluc\'{\i}a.


\end{document}